# Graphene nano-origami using Scanning Tunneling Microscopy

Kim Akius, Jan van Ruitenbeek[1]

Huygens-Kamerlingh Onnes Laboratory, Leiden University, Niels Bohrweg 2, 2333 CA Leiden, The Netherlands

## Abstract

Graphene nanopatterning on highly oriented pyrolytic graphite (HOPG) has previously been shown using both atomic force microscopy (AFM) and scanning tunneling microscopy (STM) based techniques, including folding of intrinsic step edges and electrochemical etching or cutting in graphene layers. We here report development of a new technique, combining electrochemical etching and folding using mechanical STM manipulation. Our technique shows unprecedented control of nanoscale folding of graphene in multiple folding steps and could be a route towards STM study of magic angle graphene with *in-situ* fabricated samples.

## 1. Introduction

Graphene and other two-dimensional (2D) materials are hot topics of research at the moment. The range of proposed applications is enormous and spans everything from fundamental research problems such as superconductivity [1] and spintronics [2] to practical applications such as coatings and construction materials [3]. STM is a powerful tool in condensed matter and has been successful in elucidating quantum phenomena in beautifully tangible visuals such as quantum corals [4], Friedel oscillations [5], and 1-dimensional edge states [6]. We here report on a development of a new technique of folding graphite layers with unprecedented control, using STM. The developed technique could be a possible route towards STM study of magic angle graphene, for which strong electron correlations give rise to unconventional superconductivity [1]. Only very recently one report of such a study using STM has appeared [7].

STM is a popular instrument for studying the electronic properties of materials giving insights to local density of states, magnetic order and more. However, standard fabrication techniques are not always compatible with STM. For example, graphene and 2D materials are often encapsulated or sandwiched between two layers of insulators which serves to decouple the electronic properties of the graphene, as well as preserve the graphene from contamination from the surrounding environment [8]. In this contribution we present a possible resolution to this hurdle with a STM nanofabrication technique where layers of graphene are folded on a graphite substrate. This opens up possibilities for *in situ* exfoliation, or for exposing a clean graphite surface several layers below the surface. We also demonstrate how these layers can be manipulated which suggests this method might be used to fabricate very precise structures such as magic angle graphene.

Previous work, largely by Biro's group has shown that graphene can be patterned very accurately with an STM tip using a high bias in the range of 2.5-3 V [9, 10] and, while some folding is presented, the focus is on nanopatterning and not on folding or manipulation. Other work has been done on folding twist angle graphene using AFM [11]. Our work extends these methods combining patterning, folding, and manipulation. We demonstrate unprecedented control of the manipulation with multiple manipulation steps on the same flake while using the prepatterning in a new way, as a template for folding. We establish that small angle twist layers can be obtained for single layer folded graphene as evidenced by the observation of a small angle twist moiré pattern. This technique could be a path towards using STM to study bilayer graphene superconductivity.

## 2. Methods

Single- and multi-layer graphene layers were prepatterned and folded in a commercial Jeol SPM-4500A STM in ambient conditions. HOPG from the manufacturer NT-MDT was used, where both quality specifications ZYA and ZYB (with 0.3-0.5 and 0.6-1 degrees of mosaic spread [12] respectively) yielded similar results.

Any STM manipulation experiment faces challenges of preserving the tip condition and avoiding the tip becoming too contaminated or damaged. Here mechanically polished Pt/Ir STM tips from manufacturer Unisoko were used with no further preparation. Gold coated STM tips prepared from controlled indentation in gold thin films were also studied for tip preparation but gold was found to be too malleable at room temperature to perform the folding step, resulting in smearing the gold on the surface. Hand-cut tips made from pure platinum wire however also allowed to successfully fold single-layer graphene, suggesting the possibility of *in-situ* tip preparation using platinum thin films. As platinum also has catalytic properties that can be useful for cleaning graphene [13], device design using platinum leads could be a promising path forward, enabling tip preparation on the leads.

The prepatterning was performed using a STM bias in the range of 2.5-2.7 V that is kept constant as the tip moves over the graphite surface, resulting in prepatterned cuts, such as the one shown in Fig. 1a. In the examples demonstrated here the tip was always tracking a V-shape. The optimized parameters, found by first studying the influence of various parameters when forming point-like holes, roughly agree with parameters found in previous work [10] (see supplementary information). Through this systematic analysis it was found that the bias dramatically alters the hole size and depth, while the biasing time, and the current setpoint used during the biasing, had only a minor influence on the hole size. Similarly, the influence of tip movement speed was studied for shaping lines. It was found that movement speeds in the range 2-5 nm/s were most reliable at making regularly shaped cuts.

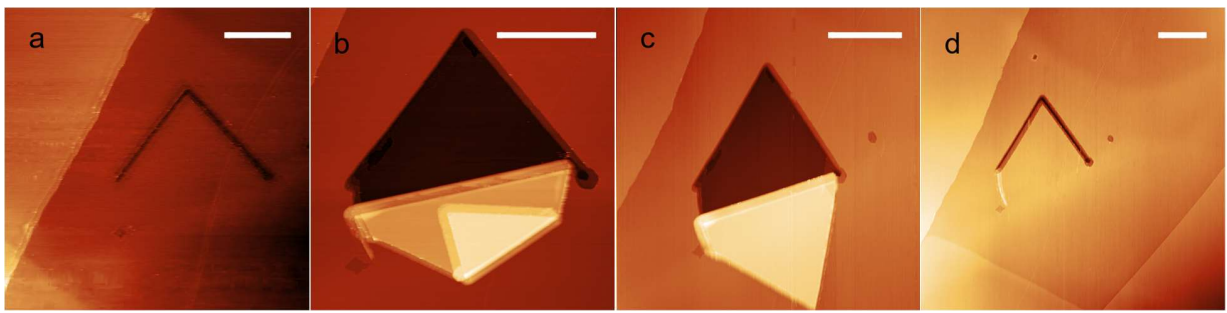

*Figure 1: Graphite (MLG) patterning and folding using STM. a.) V-groove pattern formed using high bias between sample and STM tip. b.) First folding step of few layer graphene, yielding a doubly folded region. c.) Second folding step unfolding the doubly folded region to a single fold. d.) Folding back of the same region. Note the tear on the left side, appearing as a bright line. The scalebar is 100 nm in all the images.*

After the pattern is made, the folding is performed. Here the tip is placed over the elbow in the V-groove and lowered several nm into the cut and then moved in the folding direction. If the tip is not lowered deep enough into the cut, no fold is observed and the configuration of graphene layers remains unchanged. If the cut is several monolayers deep, and the tip is pushed sufficiently deep into the cut, a multilayer graphene (MLG) fold is obtained, as in Fig. 1. After the tip is displaced in the intended direction of the fold the feedback is turned back on so that the contact between tip and sample is broken and the result of the folding operation can be imaged. If needed, as in Fig. 1b, the flake can be further manipulated to obtain a flat flake, as in Fig. 1c. The surprising robustness of this procedure is evidenced by the fact that it is possible to manipulate the same flake several times. Perhaps this is best highlighted by the flake in Fig. 1d, completely folded back into its original position, displaying a tear from the folding procedure on the left side of the cut (appearing as a bright line). The prepatterning is performed using an external Python interface to control the STM, changing STM bias, setpoint and feedback settings quickly as well as controlling the tip position. Fig. 1 illustrates what happens if this timing is not carefully controlled: the bias has stayed high a little too long, resulting in a circular shape at the end of the cut on the right side, most evident in Fig. 1b. During prepatterning, the feedback is on, but the feedback gain is low, ensuring uniformity of the etched lines, independent of any surface tilt or irregularities. After the cut the STM tip is often slightly unstable, and can usually be stabilized by performing a regular scanning, as can be seen in many of the images.

## 3. Results and discussion

Single layer graphene was folded on HOPG substrates using this method as seen in Fig. 2 which shows a sequence of intermediate images, in between manipulation steps, cutting out a predefined pattern and folding single layer graphene several hundreds of nm. Small angle moiré patterns with lattice constants of approximately 4 nm were obtained, as can be seen in Fig. 3a. Two regions display a moiré pattern, more clearly seen in Fig. 3b, both in the center of the image as well as at the bright fold in the top right. The inset of the same figure shows a Fast Fourier Transform (FFT) of a selected region in the center, obtained using the open source SPM software Gwyddion and its FFT functionality [14], displaying clear hexagonal symmetry, preserving the graphite symmetry as expected in a twist moiré. Using the relationship $D = d2\sin(\Theta/2)$ where $d = 0.25$ is the lattice constant of graphite and D is the measured lattice constant of the moiré, we obtain an angle, Θ, between the layers of 3.5 ± 0.5 degrees. After the flake was flattened out by further tip manipulation, the moiré pattern disappeared. Atomic resolution images (see supplement) analyzed with FFT suggest that the lattices are aligned after the flake is flattened, although the accuracy of judging the alignment is limited by room temperature drift.

For obtaining magic angle bilayer graphene, one needs be able to align the lattices of the bilayer with a small angle, approximately 1º. Only approximately 5% of the folds studied using our method spontaneously develops a moiré pattern. Given that it is unlikely that the folds made all by chance have ended up being in perfect registry with the underlying layer, other explanations need to be considered. Firstly, if the folded flakes are composed of multiple layers, no moiré would be observed as the first and second layer already are likely to be in registry as they are folded together. This can relatively easily be avoided by making sure the "step height" of the fold corresponds to a single layer of graphene, approximately 0.3 nm. Another influence that has been observed, is the presence of small flakes of graphite between the layers, making the layer interaction small enough to obscure a moiré. However, residues or flakes on the surface are visible on HOPG so we can exclude those images from the analysis. Controlling for these factors, another explanation is that the films easily relax into a low energy stacking or registry.

The thermal energy of the flakes or small-angle excursions during manipulation may be sufficient to ensure that the flakes can slip into the low energy stacking. For this explanation to be plausible, it would require the lateral corrugation of the surface interaction between the layers to be small enough that the energy cost of shifting the flake is small, which is known to be valid for layered van der Waals (vdW) materials such as graphite. In fact, twisted layers of graphite have so small lateral corrugation of their surface interaction that it has been reported to display supralubricity [15]. Experimental support for the relaxation into registry comes in the form of one of the moiré patterns that has been obtained in Fig. 3, which is an intermediate state of the same fold as in Fig. 2. In this intermediate state between Fig. 2c and Fig. 2d where the flake was doubly folded in some regions, moiré patterns were observed, see Fig. 3a, in a folded single layer region as well as a doubly folded (4 layer) region (bright region in top right of Fig. 3. After the flake is further manipulated, flattening the double- and triple-layer regions, resulting in the flattened flake in Fig. 2d, the moiré pattern disappears. Since graphene is known to stiffen dramatically with layer thickness [16], the double-folded region could behave as a stiffening anchor for the entire flake, pinning the flake and thereby hindering the relaxation. We have also observed that it is not possible to simply push the flake out of registry, which, instead, simply causes tears in the flake, indicating there being significant preference for the flake to sit in low energy registry. Thus, understanding this pinning mechanism, and possibly using combinations of single and multilayer regions will be necessary to control the twist angle accurately.

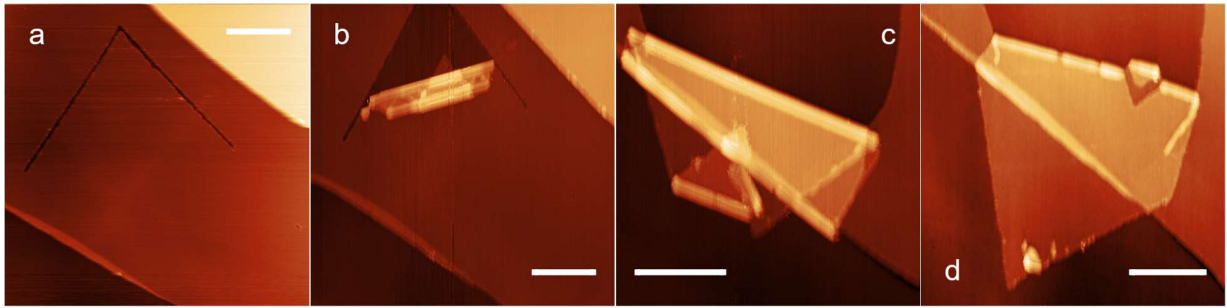

Figure 2: Folding sequence of a single layer of graphene. a.) Prepatterned V-shape in HOPG. b.) First attempt to fold a graphene flake yields a crumpled flake, half of it folded over itself. c.) Further manipulation yields a doubly folded flake where the brightest region consists of 4 layers. d.) A fully folded out single layer graphene flake, folded in several steps of STM manipulation. Note the tears on left and right side. The flake is folded all the way down over the step edge in the bottom left of 2a. The scalebar is 100 nm in all images.

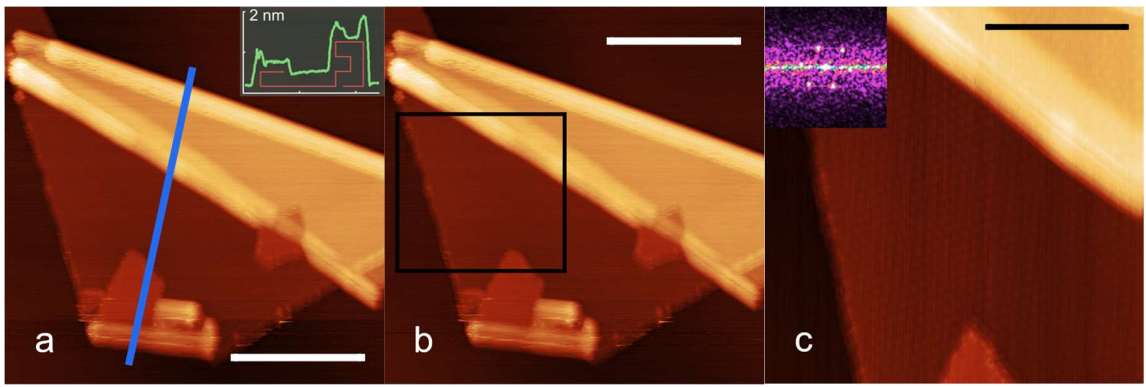

Figure 3: Folded single layer graphene flake displaying a small twist angle moiré pattern. a.) Line profile across the folded regions. Inset: The green curve is the measured line-profile along the blue line in the main panel. The red lines show a sketch of the fold configuration. b.) Overview of the regions displaying a moiré pattern; the black frame indicates the location of figure c. c.) Zoom-in of the moiré pattern regions. Inset: FFT of the moiré region of the STM image in c. The scalebar is 100 nm in a, b, and 50 nm in c.

Magic angle graphene relies on having a bilayer of graphene, not a multilayer stack such as HOPG [1]. The starting point, then, would be a single layer graphene on an insulator, for which one would need to be able to image the insulator without crashing the tip into the insulator, making it incompatible with conventional STM. This hurdle could easily be resolved using a qPlus sensor or a similar dual mode AFM/STM sensor with simultaneous capabilities to image an insulator and apply a bias between a conductive tip and sample [17]. One could transfer a single graphene layer on an insulator such as hexagonal boron nitride or silicon oxide, image using AFM mode and perform the cutting using STM mode. The study of the interesting properties of magic angle graphene require cryogenic temperatures, so that measurement cryogenic environment would be the natural next test to explore the feasibility of using these methods in real magic angle experiments. Previous work concluded that the process of cutting employed here is electrochemical and that the threshold voltage for forming cuts in UHV and inert gases is higher [18, 19]. Since cutting still occurs in UHV, this indicates that it relies on a different process in absence of airborne absorbents [20]. We plan to conduct further research aimed at controlling the pinning of the single layer graphene flakes, as well as testing under low temperature and UHV conditions. Other layered or vdW materials could also be explored, such as materials which have been proven to be very difficult to fabricate in single layers using bottom-up methods such as Bismuth(111) bilayers.

# References

[1] Cao, Y., Fatemi, V., Fang, S., Watanabe, K., Taniguchi, T., Kaxiras, E., et al., 2018. Unconventional superconductivity in magic-angle graphene superlattices. *Nature*, *556*(7699), p.43.

[2] Tombros, N., Jozsa, C., Popinciuc, M., Jonkman, H.T. and Van Wees, B.J., 2007. Electronic spin transport and spin precession in single graphene layers at room temperature. *Nature*, *448*(7153), p.571.

[3] Dimov, D., Amit, I., Gorrie, O., Barnes, M.D., Townsend, N.J., Neves, A.I., et al., 2018. Ultrahigh Performance Nanoengineered Graphene–Concrete Composites for Multifunctional Applications. *Advanced Functional Materials*, *28*(23), p.1705183.

[4] Crommie, M.F., Lutz, C.P. and Eigler, D.M., 1993. Confinement of electrons to quantum corrals on a metal surface. *Science*, *262*(5131), pp.218-220.

[5] Van Der Wielen, M.C.M.M., Van Roij, A.J.A. and Van Kempen, H., 1996. Direct Observation of Friedel Oscillations around Incorporated SiGa Dopants in GaAs by Low-Temperature Scanning Tunneling Microscopy. *Physical review letters*, *76*(7), p.1075.

[6] Drozdov, I.K., Alexandradinata, A., Jeon, S., Nadj-Perge, S., Ji, H., Cava, R.J., Bernevig, B.A. and Yazdani, A., 2014. One-dimensional topological edge states of bismuth bilayers. *Nature Physics*, *10*(9), p.664.

[7] Kerelsky, A., McGilly, L., Kennes, D.M., Xian, L., Yankowitz, M., Chen, S., et al., 2018, Magic Angle Spectroscopy, arXiv:1812.08776

[8] Dean, C.R., Young, A.F., Meric, I., Lee, C., Wang, L., Sorgenfrei, S., et al., 2010. Boron nitride substrates for high-quality graphene electronics. *Nature nanotechnology*, *5*(10), p.722.

[9] Tapasztó, L., Dobrik, G., Lambin, P. and Biró, L.P., 2008. Tailoring the atomic structure of graphene nanoribbons by scanning tunnelling microscope lithography. *Nature nanotechnology*, *3*(7), p.397.

[10] Dobrik, G., Tapasztó, L., Nemes-Incze, P., Lambin, P. and Biró, L.P., 2010. Crystallographically oriented high resolution lithography of graphene nanoribbons by STM

lithography. *physica status solidi (b)*, *247*(4), pp.896-902.

[11] Rode, J.C., Smirnov, D., Belke, C., Schmidt, H. and Haug, R.J., 2017. Twisted Bilayer Graphene: Interlayer Configuration and Magnetotransport Signatures. *Annalen der Physik*, *529*(11), p.1700025.

[12] Ohler, M., Baruchel, J., Moore, A.W., Galez, P. and Freund, A., 1997. Direct observation of mosaic blocks in highly oriented pyrolytic graphite. *Nuclear Instruments and Methods in Physics Research Section B: Beam Interactions with Materials and Atoms*, *129*(2), pp.257-260.

[13] Longchamp, J.N., Escher, C. and Fink, H.W., 2013. Ultraclean freestanding graphene by platinum-metal catalysis. *Journal of Vacuum Science & Technology B, Nanotechnology and Microelectronics: Materials, Processing, Measurement, and Phenomena*, *31*(2), p.020605.

[14] Nečas, D. and Klapetek, P., 2012. Gwyddion: an open-source software for SPM data analysis. *Open Physics*, *10*(1), pp.181-188.

[15]Liu, Z., Yang, J., Grey, F., Liu, J.Z., Liu, Y., Wang, Y., et al., 2012. Observation of microscale superlubricity in graphite. *Physical review letters*, *108*(20), p.205503.

[16] Chen, X., Yi, C. and Ke, C., 2015. Bending stiffness and interlayer shear modulus of few-layer graphene. *Applied Physics Letters*, *106*(10), p.101907.

[17] Giessibl, F.J., Hembacher, S., Herz, M., Schiller, C. and Mannhart, J., 2004. Stability considerations and implementation of cantilevers allowing dynamic force microscopy with optimal resolution: the qPlus sensor. *Nanotechnology*, *15*(2), p.S79.

[18] Albrecht, T.R., Dovek, M.M., Kirk, M.D., Lang, C.A., Quate, C.F. and Smith, D.P.E., 1989. Nanometer-scale hole formation on graphite using a scanning tunneling microscope. *Applied Physics Letters*, *55*(17), pp.1727-1729.

[19] Park, J., Kim, K.B., Park, J.Y., Choi, T. and Seo, Y., 2011. Graphite patterning in a controlled gas environment. *Nanotechnology*, *22*(33), p.335304.

[20] Kondo, S., Lutwyche, M. and Wada, Y., 1994. Nanofabrication of layered materials with the scanning tunneling microscope. *Applied surface science*, *75*(1-4), pp.39-44.